\documentclass[twocolumn,prl,superscriptaddress,showpacs,preprintnumbers,amsmath,amssymb]{revtex4}
\usepackage{graphicx}

\begin{document}

\title{Ultra-fast quenching of binary colloidal suspensions in an external magnetic field}

\author{Lahcen Assoud}
\affiliation{%
Institut f\"ur Theoretische Physik II: Weiche Materie,
Heinrich-Heine-Universit\"at
D\"usseldorf, Universit\"atsstra\ss e 1, D-40225 D\"usseldorf, Germany\\
}%
\author{Florian~Ebert}
\affiliation{%
Fachbereich f\"ur Physik, Universit\"at Konstanz,
D-78457 Konstanz, Germany
}
\author{Peter Keim}
\affiliation{%
Fachbereich f\"ur Physik, Universit\"at Konstanz,
D-78457 Konstanz, Germany
}%
\author{Ren\'e Messina}
\affiliation{%
Institut f\"ur Theoretische Physik II: Weiche Materie,
Heinrich-Heine-Universit\"at
D\"usseldorf, Universit\"atsstra\ss e 1, D-40225 D\"usseldorf, Germany\\
}%
\author{Georg~Maret}
\affiliation{%
Fachbereich f\"ur Physik, Universit\"at Konstanz,
D-78457 Konstanz, Germany
}%
\author{Hartmut~L\"owen}
\affiliation{%
Institut f\"ur Theoretische Physik II: Weiche Materie,
Heinrich-Heine-Universit\"at
D\"usseldorf, Universit\"atsstra\ss e 1, D-40225 D\"usseldorf, Germany\\
}%

\date{\today}

\begin{abstract}
An ultra-fast quench is applied to binary mixtures of superparamagnetic colloidal
particles confined at a two-dimensional water-air interface by a sudden increase  of an
external magnetic field. This quench realizes a virtually instantaneous cooling
which is impossible in molecular systems. Using 
real-space experiments, the relaxation behavior
after the quench is explored. Local crystallites
with triangular and square symmetry are formed on different
time scales and the correlation peak amplitude of the small particles
evolves nonmonotonically in time
in agreement with Brownian dynamics computer simulations.
\end{abstract}


\maketitle

Temperature quenching belongs to the key processing techniques to
produce amorphous and crystalline solids which are considerably different
from their thermodynamically stable counterparts.
Possible applications can be found in metallurgy, ceramics and semiconductor doping.
For example, quenching is used
to construct ceramic
material with a high mechanical stability \cite{She},
to design resistance devices for integrated circuits \cite{Shao},
and to improve the optoelectronic properties of semiconductors \cite{Zavada}.

Temperature quenching techniques are  most efficient if they
are performed suddenly, i.e. if the system temperature changes on a time scale
that is much shorter than a time upon which a typical particle motion is performed.
While this can be realized in computer simulations, see e.g. \cite{Yamamoto,Yakub,Arce},
it is practically impossible to be achieved  for molecular systems where the quench
is performed by a coupling to an external heat bath. There, it takes some time
until the prescribed temperature is realized in the sample. However, as
we shall show in this Letter, an ultra-fast quenching is possible  for colloidal 
particles which move much slower and are highly susceptible to external fields.

If a suspension of superparamagnetic particles is used, an external magnetic field
induces magnetic dipoles in the particles which gives rise to  dipolar
interactions between them \cite{Koenig2005}.
In equilibrium,  temperature and magnetic field strength
determine the dimensionless coupling strength between the particles such that
temperature is strictly equivalent to the inverse square
of the magnetic field strength \cite{Hoffmann,EPJE_Konstanz}. Therefore a sudden
increase of the external magnetic field corresponds to an ultra-fast  quench towards
lower temperature. It is important to note that the increase of the external magnetic field occurs
on a time scale of a few ms  much smaller than a couple of seconds 
typically  needed by a colloidal particle to diffuse over its own size. 
Thereby - unlike molecular systems -
colloidal systems can be exposed to ultra-fast temperature quenches.

In this Letter, we exploit this idea of quasi instantaneous quenching
for a binary suspension of two-dimensional
superparamagnetic colloidal
particles confined to a planar water-air interface. At high
temperatures - or equivalently at low external magnetic field
strengths - the system is weakly correlated. After an  abrupt increase of the external
magnetic field, the response of the
system and the early stage relaxation towards its new
state is  monitored by real-space imaging. For the prescribed composition of the mixture,
the equilibrium state is a crystalline  lattice with
alternating stripes of pure triangles of the majority component
and mixed squares \cite{Assoud_EPL}.
This complicated stable crystal is not reached on the time scale explored  but the system 
reveals structural heterogeneities corresponding to local metastable patches 
of crystalline order. These crystalline
zones are forming on different time scales which also gives rise to
nonmonotonic behaviour in time for structural correlation peaks of
the small particles. The real-space experiments are in agreement
with our Brownian dynamics  computer simulations. Our results can be used 
to steer the microstructure of composite materials
upon quenching and reveal the interplay between vitrification and
crystal nucleation \cite{Shintani}.

The experimental system consists of a suspension of two kinds of spherical 
super-paramagnetic colloidal particles denoted by $A$ and $B$.
Those particles are characterized  by different diameters 
($d_A=4.5\, {\rm \mu m}$, $d_B=2.8\, {\rm \mu m}$) 
and magnetic susceptibilities ($\chi_A=6.2 \times 10^{-11} {\rm \:Am^2/T},
\chi_B=6.6 \times 10^{-12} {\rm \:Am^2/T}$). 
The relative composition $X=N_B/(N_A+N_B)$ of $B$ particles is fixed at $40\%$. 
Due to their high mass density, the particles are confined
by gravity to a flat water-air interface formed by a pending water drop.
The droplet is suspended by surface tension in a top sealed cylindrical 
hole ($6 \rm \:mm$ diameter, $1 \rm \:mm$ depth) on a glass plate. 
The system can be considered as ideally two dimensional since the thermally
activated ``out of plane'' motion of the particles is 
of the order of a few tens of nanometer.
A coil produces a magnetic field $\vec H$ perpendicular to the
water-air interface inducing a magnetic moment ($\vec m_i = \chi_i \vec H$ with $i=A,B$) 
in each particle which leads to a repulsive dipole-dipole pair interaction of the form
%
\begin{equation} 
\label{eq:u_r}
u_{ij}(r)=\frac{\mu_0}{4\pi}\chi_i \chi_j H^2/r^3
\quad  (i,j=A,B).
\end{equation}
%
For this inverse power potential, at fixed composition $X$ and susceptibility ratio
$\chi_B/\chi_A$, all static quantities
depend solely \cite{Hansen_MacDonald_book} on a dimensionless interaction strength 
(or coupling constant) 
%
\begin{equation} 
\Gamma=\frac{\mu_0}{4\pi}\frac{\chi_A^2H^2}{k_BTa^3}  
\end{equation}
%
where $k_BT$ is the thermal energy at room temperature
and  $a=1/\sqrt{\rho_A}$ the average interparticle separation between $A$ particles. 
The partial $A$ particle density is set to  
$\rho_A = 1.97 \times 10^{-3} {\rm \mu m^{-2}}$ so that $a \simeq 22.5 {\rm \mu m}$.
The pair interaction is  directly controlled over a
wide range via the magnetic field. 
Making use of video microscopy, trajectories of all particles in the field of view can be recorded
over several days providing the whole phase space information.
The quench was realized upon suddenly rising the coupling $\Gamma$ 
from $1$ to $71$ at time $t=0$ by increasing the magnetic field.
The time scale of the quench is only limited by the electronics of 
power supply and was measured to be 5 ms.
This is much faster than the typical Brownian time
$\tau = d_A^2/D_A$ needed for an A particle to diffuse over its 
own distance with $D_A=0.11 {\rm \mu m^2/s}$ being the short-time diffusion 
constant for $A$ particles. The time $\tau$ is measured to be $50\,{\rm s}$.
Hence the quench can be considered to be ultra-fast. In a unquenched system 
clear deviations of purely diffusive behavior appear already at $\Gamma = 30$. 
For $\Gamma=71$ a well pronounced plateau between $t/\tau=2$ and $t/\tau=200$ is observed 
in the mean square displacement with an inflection point at $t/\tau=40$ \cite{Koenig2005}. 
Hence the quench can be considered to be ultra-fast and deep into the supercooled stage.

In parallel, we  perform  nonequilibrium Brownian dynamics (BD) computer simulations \cite{Stirner05}
of our experimental system described above. 
We employ point-like dipoles that interact following Eq. \eqref{eq:u_r} with $\chi_B/\chi_A = 10\%$ 
and $X=40\%$ in accordance to the experimental parameters. 
Knowing that the diffusion constant scales with the inverse of the radius of
a particle, $D_B$ was chosen such that
$D_B/D_A=d_A/d_B$. 
$N_A=400$  $A$ particles and $N_B=267$
$B$ particles were placed in a square box with periodic boundary conditions
applied in the two directions. 
A finite time step  $\delta t=6 \times 10^{-4}\tau$ was used.
\begin{figure}[htb]
    \begin{center}
     \includegraphics[width=8.0cm]{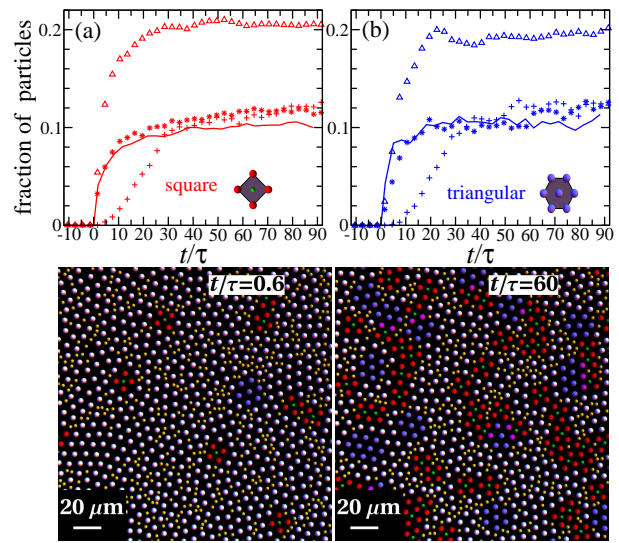}
     \caption{(a) Fraction of $B$ particles belonging to 
       a crystalline square surrounding (see inset) and
       (b) fraction of $A$ particles belonging to
       a crystalline triangular surrounding (see inset) versus reduced time
       $t/\tau$ for an ultrafast quench from  $\Gamma =1$ to $\Gamma =71$.
       The lines are experimental data while the symbols ($\ast$) are data from BD simulations.
       Two experimental snapshots for a time $t/\tau=0.6$ just after the quench (left configuration)
       and a later time  $t/\tau=60$ (right configuration) are shown.
       Big particles are shown in blue if they belong to a triangular surrounding and in red if the
       belong to a square surrounding. All other big particles are shown in white color. 
       Few big particles belonging to both
       triangular and square surroundings are shown in pink color.
       The small particles are shown in green if they belong to a square center of big particles,
       otherwise they appear in yellow.
       Also included are BD simulation data for an instantaneous "steepest descent" quench
       from  $\Gamma =1$ to $\Gamma =\infty$ ($\bigtriangleup$) and for a linear 
       increase of $\Gamma$ from $\Gamma=1$ to 
      $\Gamma=71$ on a time scale of $30\tau$ ($+$).}
    \label{SQ_TR}
     \end{center}
 \end{figure}
The early stage of the system response to the ultra-fast quench has been
observed both in real-space experiments and BD computer simulations
and was characterized by various time-dependent correlations.
Our first aim is to identify the dynamical formation of local crystallites after the quench
and detect building blocks of the underlying equilibrium 
crystal. The latter consists of alternating stripes
of pure $A$ triangles and intersecting squares of $A$ and $B$ particles \cite{Assoud_EPL}. Therefore
we have used criteria to define $A$ particles which have a pure triangular
surrounding of other $A$ particles, 
i.e.\ which are close to a cut-out of a pure triangular $A$ crystal, and,
likewise, we have identified $A$ and $B$ particles which form locally an equimolar square lattice
$S(AB)$\cite{footnote1}. The corresponding two structure elements are shown as  insets in Figure \ref{SQ_TR}. 
In detail, we associate
a triangular surrounding to an $A$ particle if the following two criteria are fulfilled simultaneously
\cite{EPJE_Konstanz}: 1) The 6-fold bond order parameter $p_6 = \sqrt{\Psi_6^*\Psi_6}$
(where $\Psi_6=\frac16 \sum_{NN}^6 \exp{{(i6\theta_{NN})}} $ with $\theta_{NN}$ denoting the
angles  of the six nearest neighbour bonds relative to a fixed reference) is larger than $0.94$.
2) The relative bond length deviation
$b_6=\frac16 \sum_{NN}^6\frac{|l_{NN}-\bar{l}|}{\bar{l}}$
where $\bar{l}$ is the average length of the six bond lengths $l_{NN}$ is smaller than $0.04$.
This double condition selects local configurations close to those of a perfect
 triangular lattice where $p_6$ is unity and $b_6$ vanishes. Likewise we define a square
surrounding around a $B$ particle by the criteria:
1) The 4-fold bond order parameter $p_4 = \sqrt{\Psi_4^*\Psi_4}$
(where $\Psi_4=\frac14 \sum_{NN}^4 \exp{{(i4\theta_{NN})}} $ with $\theta_{NN}$ denoting the bond
angles of the four nearest neighbour $AB$ bonds) is larger than $0.92$. 2) The corresponding
relative $AB$ bond length deviation $b_4$ is smaller than $0.05$.

Experimental snapshots before and after the instantaneous quench are shown in Figure \ref{SQ_TR}
with color-coded particles indicating the locations of $A$ and $B$ particles which belong to local
triangular and square clusters. If an $A$ particle has a triangular surrounding
all seven $A$ particles including the central one  with its full surrounding are shown in blue.
Conversely, if a $B$ particle has a square surrounding it is colored in green and its four
$A$ neighbours are colored in red. All particles which belong both to the blue and red class
are shown in pink.
From the snapshots of Fig.\ \ref{SQ_TR}, it is evident that the crystalline clusters
form locally and  grow as a function of time. Preferentially
triangular structures form in an area depleted from $B$ particles while square structures
nucleate around "seeds" which possess  a structure close to an underlying square. 
The resulting crystalline patches are then
fluctuating with a life-time of about $30\tau$. 
The fraction of $A$ particles with a triangular and of $B$ particles with a 
square surrounding are shown
as a function of time in Figure \ref{SQ_TR} a) and b), respectively. 
On the time scale considered
in this plot one finds an increase from almost zero before the quench to 12
percent for the triangular $A$ particles and the square $B$ particles.
 The time scales upon which triangular and square structures are formed are
different by a factor of about 2. The triangular structure is forming more rapidly than the
square one. This follows from the fact that an $AB$ square structure requires more fine-tuning
of structural correlations of both species than a triangular one which can directly
emerge in regions depleted from small particles. The number of 
pink $A$ particles 
which belong to both triangular and square surroundings is growing on a time scale
slightly longer than that for the square structure to a fraction of only 1-2 percent
far away from the equilibrium structure
where the fraction of pink particles is 2/3.
Last but not least, within the statistical error, there is  agreement
between the experimental data and our nonequilibrium BD computer simulations.

After a relaxation time of $t=10\tau$, we have further calculated the 
mean-square displacements for particles which are crystalline (i.e colored in blue, red, green) 
and non-crystalline particles during a time of $60\tau$, see Figure 1.
In both experiment and simulation, for both particles species the mean-square displacement 
of the uncolored particles is twice as large as that of the colored ones. 
This gives clear evidence for a correlation between local slow dynamics and local crystallinity.

By simulation we have finally explored a different depth and rate of the quench (see Fig.\ 1).
While a "steepest descent" quench leads to a faster growth of crystalline patches 
and to an almost doubled crystalline fraction of particles, a linear 
increase of $\Gamma$ within a time window of $30\tau$ delays their formation accordingly. 
\begin{figure}[htb]
     \includegraphics[width=8cm]{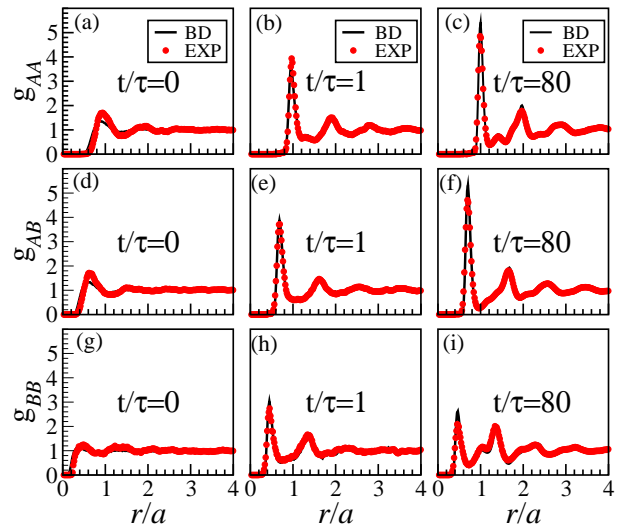}
     \caption{Partial pair distribution functions $g_{AA}(r)$,$g_{AB}(r)$ and  $g_{BB}(r)$ of
            A- and B-particles versus reduced distance $r/a$ at three different reduced times
            (a),(d),(g) $t/\tau=0$; (b),(e),(h) $t/\tau=1$; (c),(f),(i) $t/\tau=80$. BD
             results (solid lines) are compared to experimental data (symbols).}
    \label{gr}
  \end{figure}

\begin{figure}[htb]
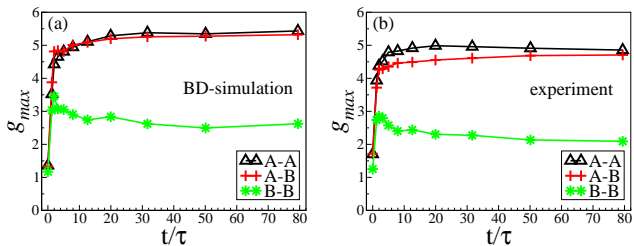

     \includegraphics[width=4.0cm]{fig3a.eps}
     \hspace{0.1cm}
     \includegraphics[width=4.0cm]{fig3b.eps}
     \caption{Amplitude $g_{max}$ of the first correlation peak in the partial pair
           distribution function $g_{AA}(r)$ (triangles), $g_{AB}(r)$ (crosses),
           $g_{BB}(r)$ (stars) as a function of reduced time $t/\tau$.
           Note the non-monotonicity in the amplitude of $g_{BB}(r)$. (a) Brownian dynamics data,
           (b) experiments.}
    \label{peak}
 \end{figure}

Next we show the time evolution of the partial pair distribution
functions $g_{AA}(r)$, $g_{AB}(r)$  and  $g_{BB}(r)$ for three times $t/\tau=0,$
$1,$ $80$ in Figure \ref{gr}. First of all, there is again
agreement between simulation and experiment
While the correlations in the starting configurations before the
quench are weak, they are quickly increasing as a function of time
towards a strongly correlated glass. An averaged  square order can
be extracted from the growth of an intermediate peak in $g_{AA}(r)$
at $r=\sqrt{2} a$. This peak grows much slower than the first peak
amplitude. The growth of the first peak amplitude in
$g_{AA}(r)$ and $g_{AB}(r)$ are monotonic in time, whereas there is a non-monotonicity in
that of $g_{BB}(r)$. This is clearly visible in Figure \ref{peak}
where the dynamical evolution of the  amplitudes of all three
partial pair distribution function is shown. With in the statistical accuracy, the peaks of
$g_{AA}(r)$ and $g_{AB}(r)$ grow on the same time scale in a
monotonic way, while the peak of  $g_{BB}(r)$ overshoots its final equilibrium
limit, both in experiment and simulation. \cite{nonmo}. 
We explain this striking
effect by a two-stage relaxation process of the small particles
which are first excluded from the triangular crystallites regions of
the big ones. Concomitantly they show a strong correlation since
they are compressed until the total system relaxes back to a state
where the small particle optimize their correlations in the network
dictated by the big ones.

\begin{figure}[htb]
     \includegraphics[width=8cm,angle=0]{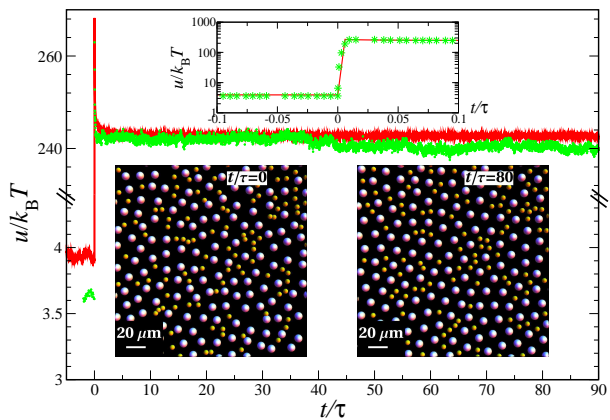}
     \caption{Time evolution of the potential energy $u$ per particle in units of $k_BT$
       versus reduced time $t/\tau$.
       Brownian dynamics simulation data are shown in red, experimental data in green.
       For a better comparison the jump after the quench is truncated (see the cut in the y-axis).
       The inset is an expanded view. Corresponding experimental snapshots are shown before 
       the quench at $t/\tau=0^-$ on the
       left side and after the quench at $t/\tau=80$ on the right side. Big particles are shown
       in white while small particles are presented as yellow spheres.}
    \label{energy}
 \end{figure}

The energetic optimization to the final state is shown in Figure \ref{energy}
revealing agreement between experiment and simulation within the statistical inaccuracies.
There is a huge jump in the potential energy per particle immediately after the quench which then
relaxes quickly towards a transient state. Then, on a second time scale, we find a slower 
relaxation process accompanied by a structural ordering as can be seen in the corresponding
experimental snapshots right before and after the instantaneous quench (see  inset of fig.  \ref{energy}).

In conclusion, we have realized an ultra-fast quench in two-dimensional colloidal mixtures
from a weakly interacting to a strongly interacting situation by a sudden increase of
an external magnetic field which controls the interparticle repulsion. The system
spontaneously relaxes  by exhibiting structural inhomogeneities
which reflect the underlying stable crystal
and correlate with slower regions in the dynamics.
The experimental data are in good agreement with Brownian dynamics computer simulations.

Our real-space characterization allows to identify the pathways of relaxation 
into a quenched glassy phase. As the quench is ultra-fast, the dynamical pathways are not blurred by
an additional time-scale from the quench history.
The built-up and the fluctuations of local crystallites after the quench can be directly
followed. Therefore our analysis can help in a more fundamental way to understand
the interplay between
vitrification and crystallization \cite{Shintani}. In fact, the structural heterogeneities 
detected here  give a considerable weight to concomitant dynamical heterogeneities 
\cite{Sperl,Onuki} and therefore represent a  relevant contribution to the
dynamically heterogeneity of glasses \cite{Harrowell,Tanaka}. Finally, our results may also be useful to
identify pathways of defect annealing in the crystalline phase
\cite{defect,Pertsinidis_Nature_2001,Pertsinidis_PRL_2001}.

Acknowledgments: We thank Patrick Dillmann for helpful discussions.
This work was supported by the DFG (projects C2
and C3 of SFB TR6 and SPP 1296).

\end{document}